\documentclass{aa}  

\usepackage{graphicx}
\usepackage{array}    
\usepackage{amssymb}    
\usepackage{url}
\usepackage{color}
\usepackage{multirow}
\usepackage{makecell}

\usepackage{adjustbox}

\usepackage{txfonts}

\def\tef{$T_{\rm eff}$}

\begin{document} 

   \title{Periodic variable A-F spectral type stars in the southern TESS continuous viewing zone\thanks{Table 4 is only available in electronic form at the CDS via anonymous ftp to cdsarc.u-strasbg.fr (130.79.128.5) or via http://cdsweb.u-strasbg.fr/cgi-bin/qcat?J/A+A/}}

   \subtitle{I. Identification and classification}

   \author{M. Skarka
          \inst{1,2,3}
          \and
          Z. Henzl\inst{3,4}
          }
   \institute{Astronomical Institute of the Czech Academy of Sciences, Fri\v{c}ova 298, CZ-25165 Ond\v{r}ejov, Czech Republic\\
              \email{skarka@asu.cas.cz}
         \and
             Department of Theoretical Physics and Astrophysics, Masaryk University, Kotl\'{a}\v{r}sk\'{a} 2, CZ-61137 Brno, Czech Republic
         \and
            Variable Star and Exoplanet Section of the Czech Astronomical Society, Vset\'{i}nsk\'{a} 941/78, 757 01 Vala\v{s}sk\'{e}, Mezi\v{r}\'{i}\v{c}\'{i}, Czech Republic
         \and
            Hv\v{e}zd\'{a}rna Jaroslava Trnky ve Slan\'{e}m, Nosa\v{c}ick\'{a} 1713, Slan\'{y} 1, CZ-27401 Slan\'{y}, Czech Republic
             }

   \date{Received May 16, 2022; accepted July 21, 2022}

 
  \abstract
   {
   } 
   {Our primary objective is to accurately identify and classify the variability of A-F stars in the southern continuous viewing zone of the TESS satellite. The brightness limit was set to 10\,mag to ensure the utmost reliability of our results and allow for spectroscopic follow-up observations using small telescopes. We aim to compare our findings with existing catalogues of variable stars. 
   }
   {The light curves from TESS and their Fourier transform were used to manually classify stars in our sample. Cross-matching with other catalogues was performed to identify contaminants and false positives.}
   {We have identified 1171 variable stars (51\,\% of the sample). Among these variable stars, 67\,\% have clear classifications, which includes $\delta$\,Sct and $\gamma$\,Dor pulsating stars and their hybrids, rotationally variables, and eclipsing binaries. We have provided examples of the typical representatives of variable stars and discussed the ambiguous cases. We found 20 pairs of stars with the same frequencies and identified the correct source of the variations. Additionally, we found that the variations in 12 other stars are caused by the contamination with the light of faint nearby large-amplitude variable stars. To compare our sample with other variable star catalogues, we have defined two parameters reflecting the agreement in identification of variable stars and their classification. This comparison reveals intriguing disagreements in classification ranging from 52\,\% to 100\,\%. However, if we assume that stars without specific types are only marked as variable, then the agreement is relatively good, ranging from 57\,\% to 85\,\% (disagreement 15-43\,\%). We have demonstrated that the TESS classification is superior to the classification based on other photometric surveys.} 
   {The classification of stellar variability is complex and requires careful consideration. Caution should be exercised when using catalogue classifications.}

    \maketitle
%

\section{Introduction}

In order to study the physical origin of stellar variability, it is necessary to investigate groups of stars that have similar observational characteristics. This means that we need clear observational criteria to accurately classify the type of variability being observed. 
Since the stellar variability is most prominent in the brightness variations, the usual and natural way of the classification is based on the shape and amplitude of the light curves and typical time scales of the variations \citep{Samus2009,Watson2006}. However, even for stars with well-defined classification parameter spaces, such as those with large amplitudes of brightness variations and typical light-curve shapes like eclipsing binaries or RR Lyrae stars, classification can be a challenging task. This calls for the need of visual inspection and utilization of additional information and various analytical methods \citep[e.g.][]{Soszynski2016,Soszynski2019,Prsa2022,Iwanek2022}.

Classification of stars that exhibit low-amplitude variations caused by high-order radial and non-radial pulsations, spots of various types, deformation of the spherical shape of stars, and stellar activity is a challenging task. This is because these effects have similar amplitudes and time scales, thereby making it difficult to distinguish between them. The situation is further complicated by a combination of different mechanisms that can be present simultaneously, especially for stars with intermediate masses of $1-3$\,M$_{\odot}$ (F-A spectral-type stars). 

Moreover, the classification of stars becomes more complicated with the availability of ultra-precise space observations from the {\it Kepler} \citep{Borucki2010} and TESS satellites \citep{Ricker2015}. These observations have revealed new phenomena with similar amplitudes as the intrinsic stellar variations, which are related to the contamination with nearby stars, instrumental effects, and trends induced into the observations by the data reduction procedures. Especially in the mmag and sub-mmag regime, the human supervision can be biased and subjective, making even manual classification at least partially questionable.

In some cases, the frequency limits that are commonly accepted for classifying stars by their variability are not firmly established or cannot be generally applied. For instance, \citet{Balona2022} argues that there is no need for the roAp class \citep[rapidly oscillating Ap stars,][]{Kurtz1982} because there is no frequency limit that differentiates roAp and $\delta$\,Sct (DSCT) stars. Similarly, the commonly accepted limit of 5\,c/d \citep{Grigahcene2010} that differentiates $\gamma$\,Dor (GDOR) and DSCT stars can be questionable since DSCT frequencies can appear below this limit and GDOR frequencies can be above 5\,c/d \citep{Grigahcene2010}.

Lastly, our limited understanding of the physics behind stellar variability does not allow for precise constraints on the physical limits of different physical mechanisms. These limits, in addition, cannot be generalized and are different for every particular star. For example, the position of the empirical DSCT instability strip (IS) borders is different than the theoretical limits, and there are non-pulsating stars inside the IS and pulsating stars outside this region \citep[see figures in e.g.][]{Murphy2019,Antoci2019}.

The identification and classification of variable stars is a complex task, as there are many factors that can affect the accuracy of automatic classification using modern computational methods \citep[e.g.][]{Audenaert2021,Audenaert2022}. Additional information, such as multicolour photometric, spectroscopic, and astrometric observations, are often needed to ensure reliable classification. In a recent study by \citet[][hereafter S22]{Skarka2022}, they discussed the challenges of identifying and classifying bright A-F stars in the northern TESS continuous viewing zone (CVZ) based on TESS photometric data. We proposed a conservative classification scheme and compared the results with previous studies. 

This paper presents a similar analysis for A-F stars in the southern TESS CVZ, using the same criteria and methods as the previous study. We provide conservative samples of variable stars that are suitable for follow-up spectroscopic observations with small telescopes and that can serve as a starting point for further detailed studies.

\section{Sample selection and data retrieval}\label{Sect:Sample}

We used the same methodology as in S22. Initially, we selected a sample of 337798 point sources located within a 15-degree circle around the southern ecliptic pole (RA=6$^{\rm h}$00$^{\rm m}$00$^{\rm s}$, Dec.=-66$^{\circ}$33'00'') from the TESS Input Catalogue (TIC) v8.0 \citep{Stassun2019}. From this sample, we filtered out duplicate stars and selected only 2302 stars with a record in the SIMBAD database \citep{Wenger2000} that have a temperature range of 6\,000$<$\tef$<$10\,000\,K and a brighter than 10\,mag in Johnson $V$. The temperature distribution of the stars is shown in Tab.~\ref{Tab:TefStats}. Most of the stars are fainter than 8\,mag (2043 stars), with only 17 stars being brighter than 6\,mag.

\begin{table}
\caption{Number of stars in particular temperature ranges (based on temperatures from TIC \citep{Stassun2019}).}             
\label{Tab:TefStats}      
\centering                          
\begin{tabular}{c c}        
\hline\hline                 
$T_{\rm eff}$ (K) & N \\    
\hline                        
$6000-7000$ & 1308 \\      
$7000-8000$ & 603 \\ 
$8000-9000$ & 222 \\ 
$9000-10000$ & 169 \\  \hline \hline
\end{tabular}
\end{table}

We obtained all available data, which was processed by the TESS Science Processing Operations Center \citep[SPOC;][]{Jenkins2016} and the quick-look pipeline \citep[QLP;][]{Huang2020a,Huang2020b}, using the \textsc{Lightkurve} software \citep{Lightkurve2018,Barentsen2020}. We extracted the pre-search data conditioning simple aperture photometry (PDCSAP) flux with long-term trends removed \citep[][]{Twicken2010} and transformed the normalized flux to magnitudes. We did not apply any additional de-trending or data filtering to preserve possible variability.

However, it was not possible to download data from all observed sectors for some stars using \textsc{Lightkurve}. Therefore, for some stars, only one sector was available despite being observed in multiple sectors. We analyzed data from Cycles 1 (sectors 1-13) and 3 (sectors 27-39) in this paper. The SPOC routine provided 2-minute and 20-second cadence data (SC) for cycles 1 and 3, respectively, and 30- and 10-minute cadence data (LC) for cycles 1 and 3, respectively. The QLP routine only provided LC data.

The Table~\ref{Tab:Points_Stats} presents an overview of the available data sets in different routines and their statistics. It is clear from the table that data for all the sample stars were available in QLP, while SC SPOC data are missing for more than 300 stars. The data from more than 20 sectors are available for most of the stars in SPOC LC and QLP routines, whereas about half of the stars have SC data only in 13 sectors. Additionally, the LC routines generally provide a larger time span than is available for the SC data.

\begin{table}
\caption{Statistics of the analysed data sets.}             
\label{Tab:Points_Stats}      
\centering                          
\begin{tabular}{l c c c}        
\hline\hline                 
& SC SPOC & LC SPOC & QLP \\    
\hline                        
Stars & 1978 & 2249 & 2302 \\
med Pts & 202441 & 48078 & 53095 \\
Sect  & 1-26 & 1-26 & 4-26\\ 
med Sect & 13 & 23 & 23 \\ 
TS (d) & 22-1064 & 24-1064 & 296-1064 \\
med TS (d) & 901 & 1064 & 1064 \\
\hline \hline                                   
\end{tabular}
\tablefoot{Column ‘Stars’ gives the number of stars, ‘med Pts’ gives the median of points, ‘Sect’ gives the range of available sectors, ‘med Sect’ gives the median of available sectors, ‘TS’ and ‘med TS’ are the time span of the data and its median, respectively.}
\end{table}

\section{Classification of the variable stars}\label{Sect:Identification}

We followed the same approach as discussed in S22, mainly in their sects. 3 and 4. The QLP data products have improved since the publication of S22 and are of comparable quality to the SPOC LC data. However, we used QLP data only when no SPOC data were available. The LC SPOC and QLP data contain various artificial signals and they provide only integrated brightness over a longer period, significantly reducing the amplitude of fast variations. Therefore, we used SPOC SC data whenever possible for the analysis. We used SPOC LC and QLP data only when there was a lack of SC data or when we needed to complement SC data with data having a longer time base.

We carefully examined each light curve, analyzing the most prominent frequency in the frequency spectrum, and the overall characteristics of the frequency peaks, including their position, content, and morphology. Our classification system is based on the criteria described in detail in tables 2-4 in S22. For a better orientation among the variability types used in this work, we give their basic description in Table~\ref{Tab:VarTypes}. 

\begin{table*}
\tiny
\caption{Description of variability types used in this work together with the basic light-curve and frequency spectra characteristics.}             
\label{Tab:VarTypes}     
\centering                          
\begin{tabular}{l c c} 
\hline\hline  
\multicolumn{3}{c}{Binarity} \\ \hline
ECL (eclipsing binary of unclear type) & Apparent minima & Harmonics of the basic frequency\\
EA, EP (Algol type, exoplanet) & Constant light in maximum, clearly defined minima & Many well defined harmonics of the basic frequency \\ 
EB ($\beta$ Lyr type) & Variation in maximum light, well defined minima & Many well defined harmonics of the basic frequency \\
EW (W UMa type) & Smooth brightness variations & Well defined harmonics of the basic frequency \\
ELL (ellipsoidal variables) & Regular, smooth brightness variations & One or two harmonics of the basic frequency\\ \hline
\multicolumn{3}{c}{Rotation} \\ \hline
ROTM (magnetic rotators) & Regular, smooth brightness variations & A few harmonics of the basic frequency\\
ROTS (Solar-like rotator) & Semi-regular variations & Unresolved peaks close to harmonics of the basic frequency\\
ROT (non-specified rotator) & Repeating stable features & A few harmonics of the basic frequency\\ \hline
\multicolumn{3}{c}{Pulsations} \\ \hline
RRL, DCEP (RR Lyr, Cepheids) & Fast rise to maximum, large amplitude & A few harmonics of the basic frequency\\
GDOR ($\gamma$\,Dor, g-mode pulsators) & (ir)regular variations, beating, bumps & two and more independent peaks below 5\,c/d \\
DSCT ($\delta$\,Sct, p-mode pulsators) & (ir)regular variations, beating, bumps & two and more independent peaks above 5\,c/d \\
\hline \hline                                   
\end{tabular}
\end{table*}

\begin{figure}
\centering
\includegraphics[width=0.49\textwidth]{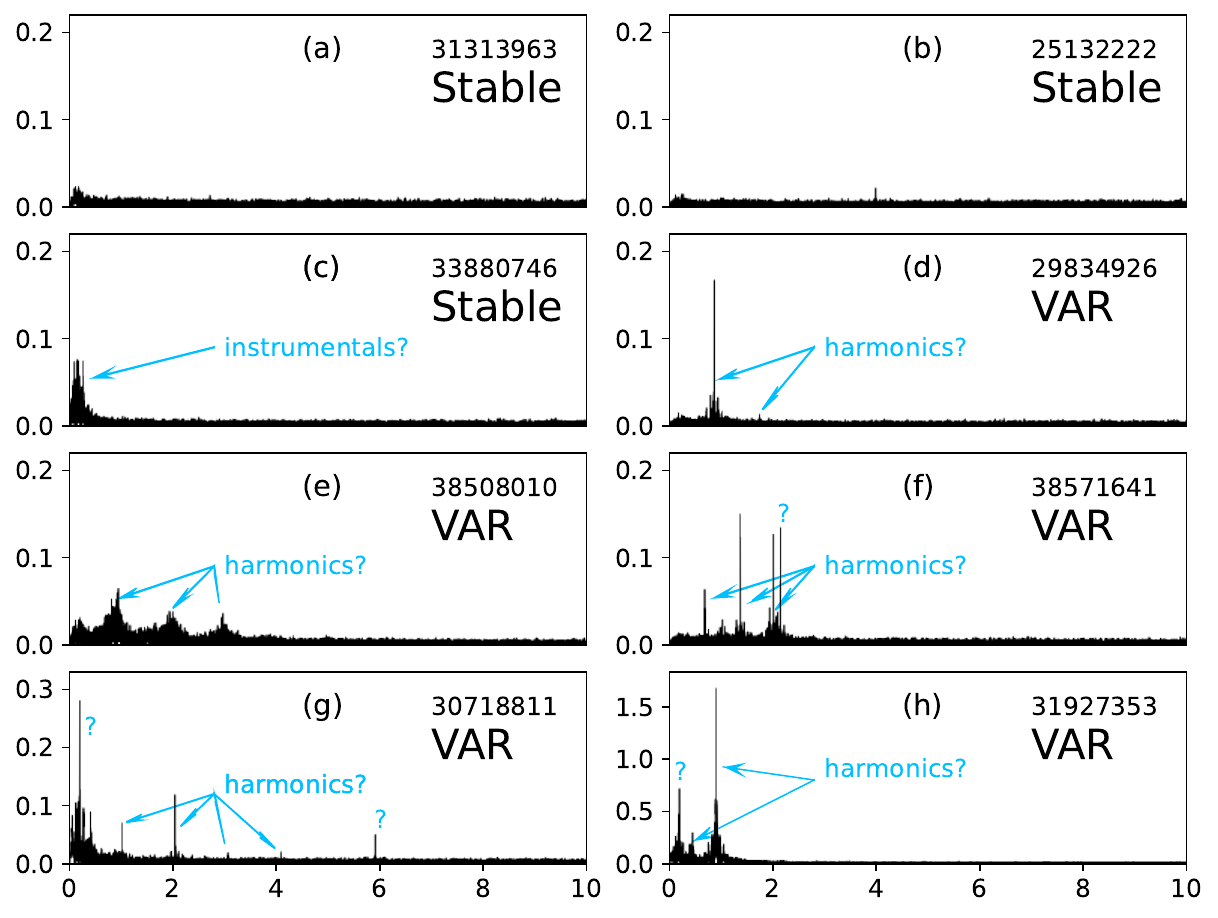}
\caption{Examples of frequency spectra of stable stars and stars marked as VAR with unclear classification. The horizontal axis shows frequency in c/d, while the vertical axis shows brightness in mmag. The numbers given in the labels are the TIC numbers.
}
\label{Fig:VAR}
\end{figure}

In cases where it was difficult to determine the variability type, but the star was clearly variable, we designated it as ‘VAR’. These stars typically exhibit probable harmonics (which can include eclipsing binaries, ROT, ROTS, ELL, ROTM, RRL, etc.) and additional peaks with unclear nature (panels d-h in Fig.~\ref{Fig:VAR}). Some VAR stars display a single peak or low-frequency (0-0.5\,c/d) and low-amplitude ($<0.1$\,mmag) peaks that may be of instrumental origin. Some of the stars classified as STABLE (panels a-c in Fig.~\ref{Fig:VAR}) can be VAR in fact, but the decision about the variability and/or the true source of variability was impossible either due to the small amplitude of the variations (mostly apparent only in the Fourier transform) or due to number of contaminants. The complete classification is provided in Table~\ref{Tab:Main}.

\begin{table*}
\small
\caption{Classification of stars in our sample.}             
\label{Tab:Main}      
\centering                          
\begin{tabular}{c c c c c c c c c c}        
\hline\hline 
TIC	&	RAJ2000	&	DEJ2000	&	CR	&	Blends	&	Type	&	$V$ (mag)	&	$T_{\rm eff}$ (K)	&	$L$ (L$_{\odot}$)	\\ \hline
25078674	&	60.18200711	&	-71.16679809	&	0.0002	&	0	&	VAR	&	6.60	&	9150	&	26.3	\\
25117240	&	61.65684312	&	-69.26155744	&	0.3208	&	2	&	GDOR	&	9.40	&	6743	&	15.4	\\
25117273	&	61.38801709	&	-69.40047833	&	0.0005	&	0	&	DSCT+LF	&	9.52	&	7175	&	10.4	\\
25117406	&	61.51260808	&	-69.87130832	&	0.0006	&	0	&	DSCT+GDOR	&	7.45	&	7231	&	36.7	\\
25117492	&	61.58357233	&	-70.24478618	&	0.0015	&	1	&		&	9.80	&	8197	&	17.3	\\
25118797	&	61.93961418	&	-69.05238972	&	0.5487	&	1	&		&	9.86	&	6122	&	3.4	\\
25132222	&	62.00132424	&	-68.36099381	&	0.0105	&	0	&		&	9.73	&	6064	&	3.3	\\
25132314	&	62.12780124	&	-68.70223768	&	0.0006	&	0	&	GDOR	&	8.38	&	6867	&	7.1	\\
25133052	&	62.13967593	&	-71.63984666	&	0.0016	&	2	&		&	8.64	&	6315	&	2.7	\\
25153007	&	62.8648749	&	-70.73842226	&	0.0024	&	0	&	ROT	&	9.22	&	7109	&	5.3	\\
...	&	...	&	...	&	...	&	...	&	...	&	...	&	...	&	...	\\
\hline \hline                                   
\end{tabular}
\tablefoot{The full table that is available at CDS. The ‘CR’ column gives the contamination ratio \citep{Paegert2021}, ‘Blends’ give the number of stars with magnitude difference less than 5\,mag from the target stars, values of $V$, $T_{\rm eff}$ and $L$ are taken from \citet{Stassun2019}.}
\end{table*}

\subsection{Stars with well defined equidistant peaks}\label{Subsect:Rotation}

Periodic variations associated with rotation (classes ROT, ROTM, ROTS), orbital motion (EA, EB, EW, ELL), or non-sinusoidal pulsations (e.g., DCEP, RRL) produce peaks at the fundamental frequency and its harmonics, resulting in an equidistant pattern in the frequency spectrum. This pattern is easy to recognize. However, the classification of these variations is not straightforward.

\begin{figure}
\centering
\includegraphics[width=0.49\textwidth]{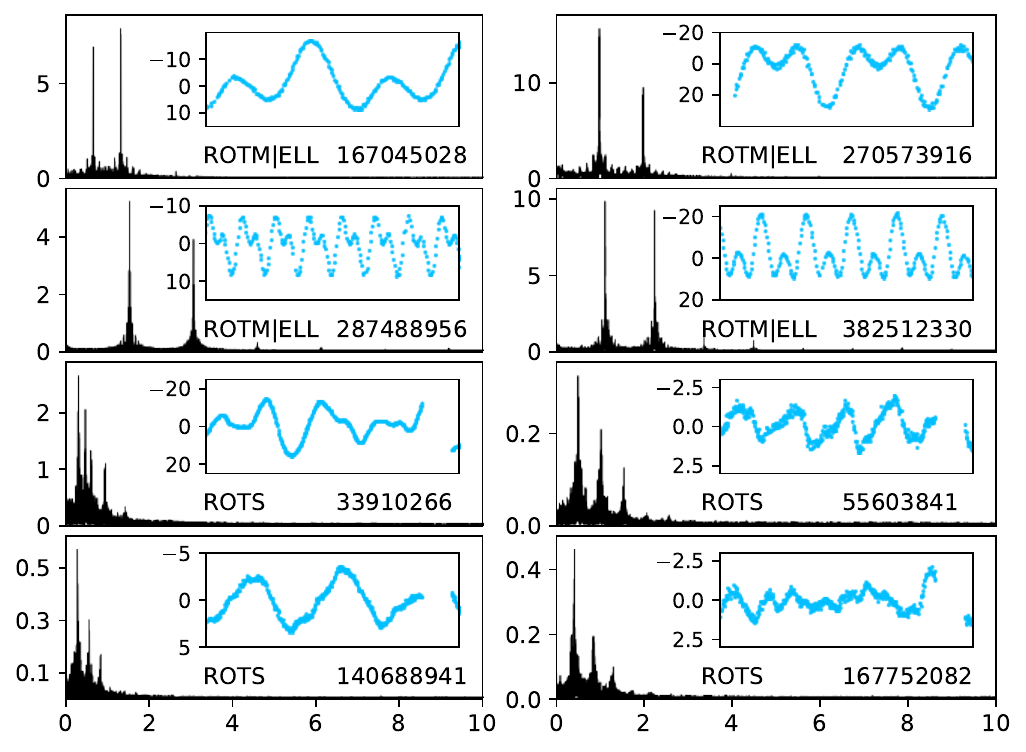}
\caption{Examples of rotationally variable stars. The horizontal axis shows frequency in c/d, while the vertical axis shows brightness in mmag. The insets show the data in 5 days (four upper panels) and 10 days, respectively. The numbers given in the labels are the TIC numbers.
}
\label{Fig:ROT}
\end{figure}

Identifying and classifying eclipsing binaries is usually an easy task due to prominent eclipses. However, the binary stars can be confused with other possible classes if the eclipses are shallow and/or deformed, and not distinct. Distinguishing non-eclipsing binary stars with tidally deformed components (ellipsoidal variables and contact binaries) from spotted chemically peculiar stars (class ‘ROTM’) based on a single-channel TESS photometry can be an extremely difficult task, as discussed in S22. Both classes show a dominant frequency and usually one well-defined harmonic in the frequency spectrum (see the four upper panels of Fig.~\ref{Fig:ROT}). The contamination of ELL stars with ROTM stars is likely not too high because the incidence rate of the spotted chemically peculiar stars of Ap type is small, only about 10\,\% among the A-spectral type stars \citep{Sikora2019a}.

The light curves of ELL (ellipsoidal) stars can be distorted by various effects like Doppler beaming, reflection effects, and spots \citep{Faigler2011}. Therefore, we decided to use a more conservative approach and classify such stars as ‘ROTM|ELL’, instead of distinguishing between ROTM (rotational modulation) and ELL. In the example shown in Fig.~\ref{Fig:ROT}, only TIC~270573916 is an ELL star with high probability. To make a final decision between ROTM and ELL classes, spectroscopic observations are necessary. In a study by \citet{Green2023}, only 50\% of 97 suspected ellipsoidal variables showed clear radial velocity variations upon spectroscopic observation, indicating possible contamination with spotted stars. We have also included stars that exhibit apparent simple variability with a single peak in the frequency spectrum to the ROTM|ELL class.

It is important to note that there can be confusion between rotation and pulsations in stars that exhibit temperature spots similar to our Sun. These stars belong to a category called ‘ROTS’ and their frequency spectra can be similar to GDOR pulsators. Active stars of the solar-like type show groups of peaks around harmonics with rotational frequency due to differential rotation. In Fig.~\ref{Fig:ROT}, the four bottom panels show typical examples of such stars with TIC~33910266 being a bit questionable, showing five harmonics with gaps at 4$f_{\rm rot}$ and $6f_{\rm rot}$.

Stars that exhibit sharp, isolated peaks and their harmonics in the frequency spectra, which cannot be classified accurately, are categorized as ‘ROT’. These stars can be of various types such as ELL, ROTM, ROTS, RRL, DCEP, eclipsing binaries, and others that display equidistant frequency peaks. Therefore, the classification of ‘ROT’ stars should be considered only as preliminary and rough similarly to stars in the VAR class. The frequency spectra of VAR stars shown in panels f-h in Fig.~\ref{Fig:VAR} can be considered as belonging to the ROT class.

Rotationally variable stars can be false positives because their light can be contaminated with the light of far-away-located large-amplitude pulsating stars (for example, RR Lyr and Cepheids) or eclipsing binaries. Such stars and systems can have periods of variations similar to rotational periods of A-F stars and show harmonics in their frequency spectra. In cases of contamination, the amplitude and number of detectable harmonics are strongly reduced. Examples of the contaminants are discussed in Sect.~\ref{SubSect:Contamination}.

\subsection{Pulsating stars}\label{Subsect:GDOR}

Assigning a star with GDOR (gravity-mode, g-mode pulsators), DSCT (pressure-mode, p-mode pulsators), GDOR+DSCT and DSCT+GDOR (hybrid pulsators) classes can also be challenging task. The first difficulty arises with the definition of the GDOR class, specifically concerning the frequency content and pattern in the frequency spectra. The gravity-mode pulsations of GDOR stars are excited by the convective flux blocking mechanism \citep{Guzik2000,Dupret2005}. This type of pulsation does not produce equally spaced structures in the frequency domain (harmonics are possible). The GDOR frequencies are usually between 0.3 and 3\,c/d \citep{Kaye1999,Henry2011} with the most likely position being approximately 1.2-1.3\,c/d \citep[see fig. 1 in][]{Tkachenko2013}. However, when shifted by the rotation, the frequencies can reach up to 5 or even more cycles per day \citep{Grigahcene2010}. The usually accepted limit to distinguish between g-mode and p-mode pulsators is 5\,c/d \citep{Uytterhoeven2011}, but other limits, such as 4\,c/d \citep{Antoci2019}, are also being utilised.

\begin{figure}
\centering
\includegraphics[width=0.49\textwidth]{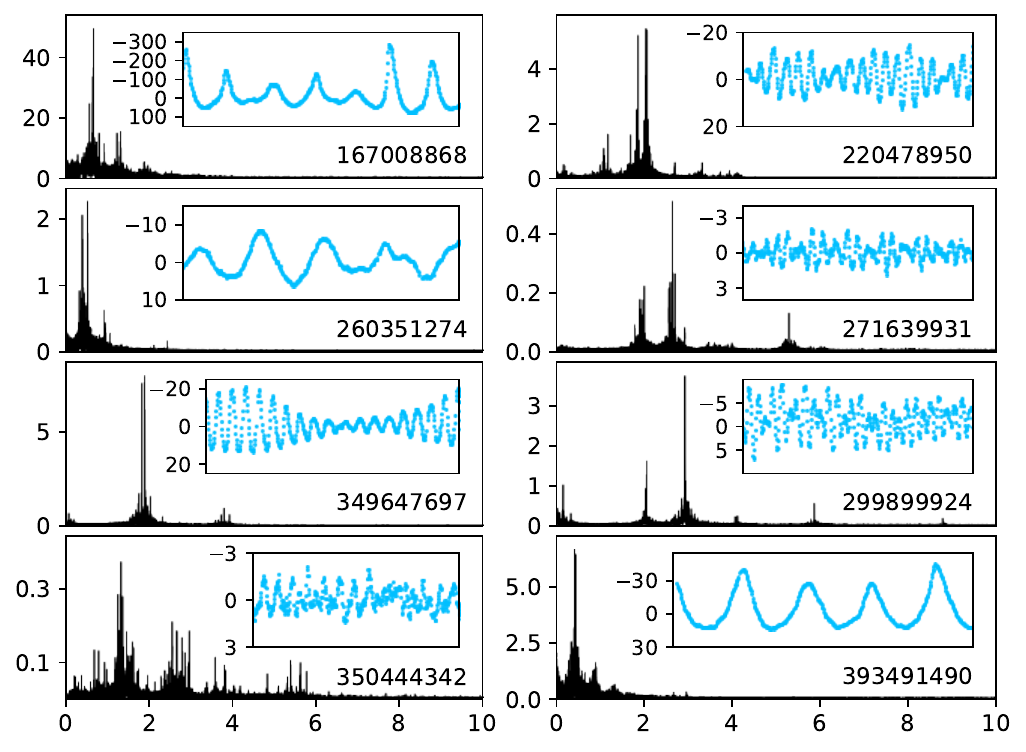}
\caption{Examples of GDOR stars. The horizontal axis shows frequency in c/d, while the vertical axis shows brightness in mmag. The insets show the data in 10 days. The numbers given in the labels are the TIC numbers.}
\label{Fig:GDOR}
\end{figure}

The frequency pattern that is characteristic for the GDOR stars is a comb of close, sometimes unresolved, peaks \citep[for example,][our Fig.~\ref{Fig:GDOR}]{Balona2011gdor,Uytterhoeven2011,Tkachenko2013} that can also be found around harmonics \citep[see fig. 3 in][and our Fig.~\ref{Fig:GDOR}]{Saio2018}. This can cause a confusion with spotted solar-type stars which is best apparent in TIC~393491490 in the bottom right panel of Fig.~\ref{Fig:GDOR}. On the other hand, the peaks can also be randomly distributed \citep[e.g.,][and TIC~350444342 in the bottom left panel of Fig.~\ref{Fig:GDOR}]{Uytterhoeven2011,Balona2011gdor}. We classified stars as GDOR when several peaks were present in the range of approx. 0.3-5\,c/d. At the same time, the frequency pattern did not comply the requirements for the rotationally variable stars (see Sect.~\ref{Subsect:Rotation}). In addition, the amplitude of the frequency peaks has to be larger than 0.1\,mmag. 

It is important to note that we do not differentiate between slowly-pulsating B stars \citep[SPB stars][]{Waelkens1998,DeCat2002} and GDOR stars as SPB stars are hotter and may contaminate our sample at high temperatures. Additionally, we do not distinguish between low- and high-amplitude GDOR stars \citep[HAGDOR,][]{Paunzen2020} because the boundary between the two is not well-defined, and amplitudes in TESS data may not be reliable due to contamination and data processing. TIC\,167008868 is an excellent candidate for HAGDOR (top left panel of Fig.~\ref{Fig:GDOR}) with one of the largest peak-to-peak amplitudes ever observed in HAGDOR (about 0.4\,mag in TESS pass band, SPOC routine).

The second important class of pulsating stars among A-F spectral type stars are DSCT pulsators. The pulsations in these low-radial-order p-mode pulsators are excited by the $\kappa$ mechanism operating in the HeII ionisation zone and by turbulent pressure \citep{Breger2000,Antoci2014}. As already mentioned, five cycles per day are usually taken as the border between GDOR and DSCT pulsations \citep{Uytterhoeven2011}. However, some authors use 4\,c/d \citep{Antoci2019}. The upper limit is also not firmly defined. \citet{Balona2022} stated that there is no sharp transition between roAp \citep{Kurtz1982} and DSCT stars and that the roAp class should be dropped. In fact, DSCT and roAp stars are different objects and such a comparison concerns only the frequency content \citep[][]{Holdsworth2024}. Nevertheless, we marked a star as a DSCT pulsator if it shows at least one strong peak above 5\,c/d. Due to the faster cadence in Cycle~3, the Nyquist reflections are not a significant problem since the Nyquist frequency is 72\,c/d which is well above the dominant frequency for most of the DSCT stars.

\begin{figure}
\centering
\includegraphics[width=0.49\textwidth]{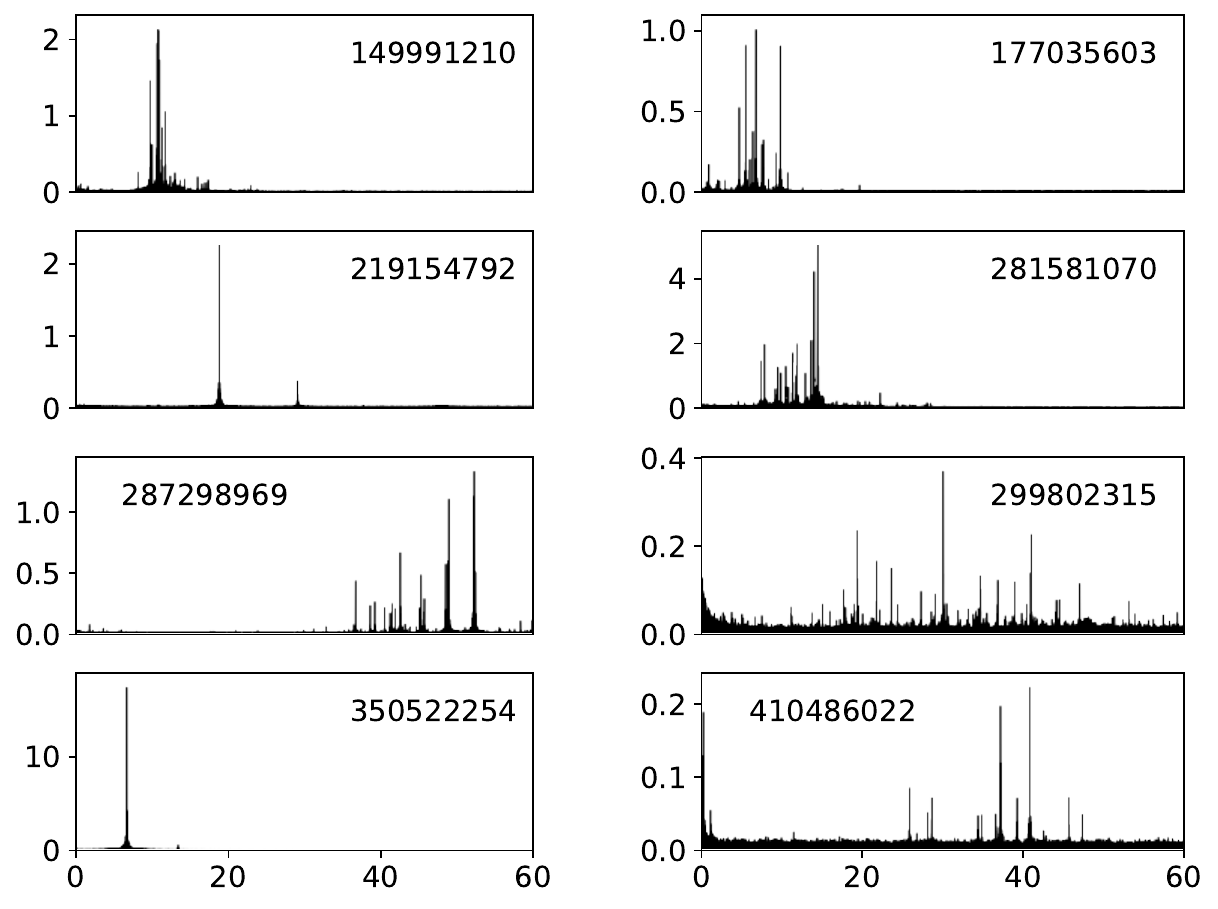}
\caption{Examples of frequency spectra of DSCT stars. The horizontal axis shows frequency in c/d, while the vertical axis shows brightness in mmag. The numbers given in the labels are the TIC numbers.}
\label{Fig:DSCT}
\end{figure}

The variety of DSCT frequency spectra has been shown many times \citep[e.g.][]{Grigahcene2010,Uytterhoeven2011}. Examples of frequency spectra of DSCT stars in our sample are in Fig.~\ref{Fig:DSCT}. We can see stars with a single dominant pulsation mode (e.g. TIC~350522254), stars with a comb of peaks (e.g. TIC~149991210) but also stars with peaks across the whole frequency range (e.g. TIC~299802315). 

The ultra-precise space observations revealed that basically, all DSCT and GDOR stars show peaks in both regimes \citep{Grigahcene2010,Uytterhoeven2011,Balona2014}. \citet{Balona2014} also showed that most of the low frequencies observed in the DSCT stars cannot be explained by the non-linear combination of the DSCT modes and by rotation. Thus, the vast majority of stars showing peaks in both DSCT and GDOR regimes, are likely hybrids. 

The only situation when it is difficult to identify hybrid pulsators is in stars where the amplitudes of the frequency peaks are different or the frequency content is poor and the frequency peaks cannot be properly interpreted. There is no commonly accepted amplitude threshold specifying that a particular star is a pure GDOR or DSCT pulsator or a hybrid. The ratio of the amplitudes in both regimes used by \citet{Uytterhoeven2011} and \citet{Bradley2015} are only empirical thresholds (factor 5-7 in amplitude ratio) not saying anything about the hybrid nature. In addition, each author uses a different amplitude limit for considering the frequencies for the classification. All of the mentioned problems add ambiguity to the classification. 

\begin{figure}
\centering
\includegraphics[width=0.49\textwidth]{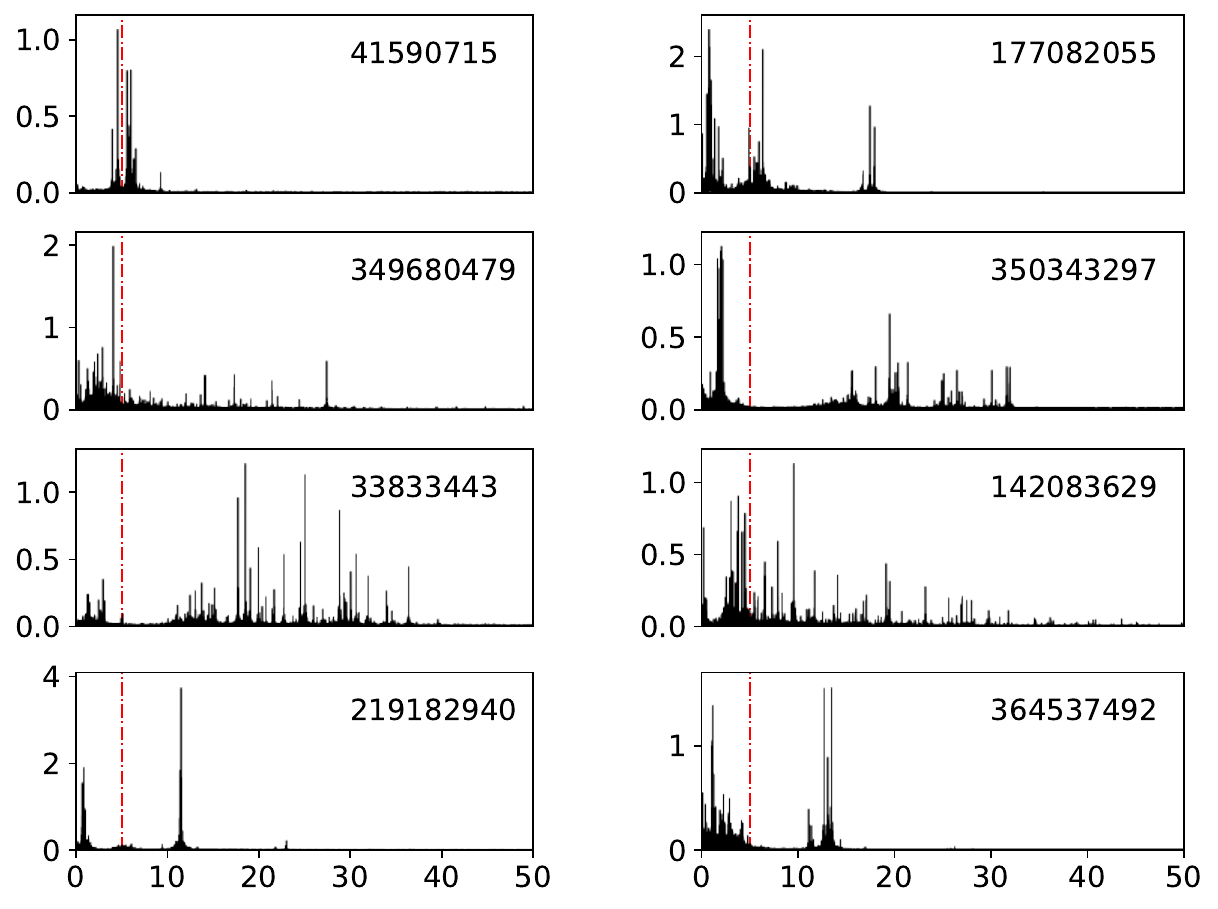}
\caption{Examples of frequency spectra of hybrid GDOR+DSCT stars. Frequency on the horizontal axis is in c/d, while brightness on the vertical axis is in mmag. The numbers given in the labels are the TIC numbers. The upper four panels show GDOR-dominating hybrids, while the four bottom panels show DSCT-dominating hybrids. The vertical red dash-dotted lines indicate 5\,c/d.}
\label{Fig:HYB}
\end{figure}

As each data set is unique, we did not adopt any amplitude threshold because even peaks in the sub-mmag regime can be significant. We cannot avoid mixing hybrids with pure DSCT or GDOR stars showing combination peaks without detailed frequency analysis, which is out of the scope of this paper. However, this will be the case in only a minority of stars. For instance, TIC~177035603 (top right-hand panel of Fig.~\ref{Fig:DSCT}) can be classified as a hybrid pulsator with similar probability as pure DSCT. Similarly, TIC~350444342 in the bottom left panel of Fig.~\ref{Fig:GDOR} can be also considered as a hybrid pulsator, not a pure GDOR star. Examples of frequency spectra of hybrid pulsators are shown in Fig.~\ref{Fig:HYB}. The least reliable classification among the shown stars is in TIC~41590715 because the peaks are grouped around 5\,c/d and the classification can be all of pure GDOR, pure DSCT, or hybrid pulsator. We give ‘DSCT+LF’ classification to more than 60\,\% of the DSCT stars showing low-amplitude peaks below 5\,c/d meaning DSCT stars with low-frequency (LF) peaks with unclear interpretation. Some of these stars can be hybrid stars, but some can also be rotationally variable.

\subsection{Contamination and identification issues}\label{SubSect:Contamination}

The southern TESS CVZ includes the Large Magellanic Cloud (LMC), where many stars are strongly affected by contamination. This is especially problematic due to the low angular resolution of TESS, which is 21 "/px \citep{Ricker2015}. Contamination can cause several issues such as distortion of the light curve, mixing of variability, and decrease of the brightness amplitude. The new data releases of {\it Gaia} catalogues \citep{GAIA2018,GaiaDR3} led to the identification of redundant entries in the TICv8.0 catalogue \citep{Stassun2019}. In TICv8.2, \citet{Paegert2021} identified "phantom" stars, which arise from diffraction spikes around bright stars, mismatches between different catalogues, and from a combination of two or more fainter stars mimicking one target. 

We have found 13 such duplicate entries in our sample by comparing TIC8.0 and TIC8.2 catalogs. The numbers of the duplicates are followed with a ‘p21’ suffix in the ‘Type’ column of Table~\ref{Tab:Main}. We also adopted the contamination ratio from \citet{Paegert2021} in column CR, which gives the ratio of contaminating flux and flux of the target star. Furthermore, we have cross-matched the positions from TIC8.0 with the {\it Gaia} DR3 catalogue \citep{GaiaDR3} within a distance of 105" (5\,px). The number of blending stars with a magnitude difference less than 5\,mag from the target star is given in the ‘Blend’ column of Table~\ref{Tab:Main}. This information can at least help to identify problematic targets. The most contaminated target is TIC\,372913472, which has 13 contaminants. Automatic procedures for identifying the true source of the variation signal \citep[e.g.,][]{Higgins2023} are not effective in our case, as they do not work well with a single frequency. Therefore, each target would require a detailed investigation of the frequency spectrum to get reliable and unambiguous results, which is beyond the scope of this paper.

We also compared the strongest identified frequencies of categorized stars and visually examined the frequency content to identify any obvious contaminants. We considered a star with a larger amplitude peak as the real variable, while the other star with the same frequency but lower amplitude was deemed a duplicate contaminated by the variable star. In Table~\ref{Tab:Main}, we provide the TIC number of the real variable (the contaminant) in the ‘Type’ column. There are 20 such cases.

\begin{figure}
\centering
\includegraphics[width=0.48\textwidth]{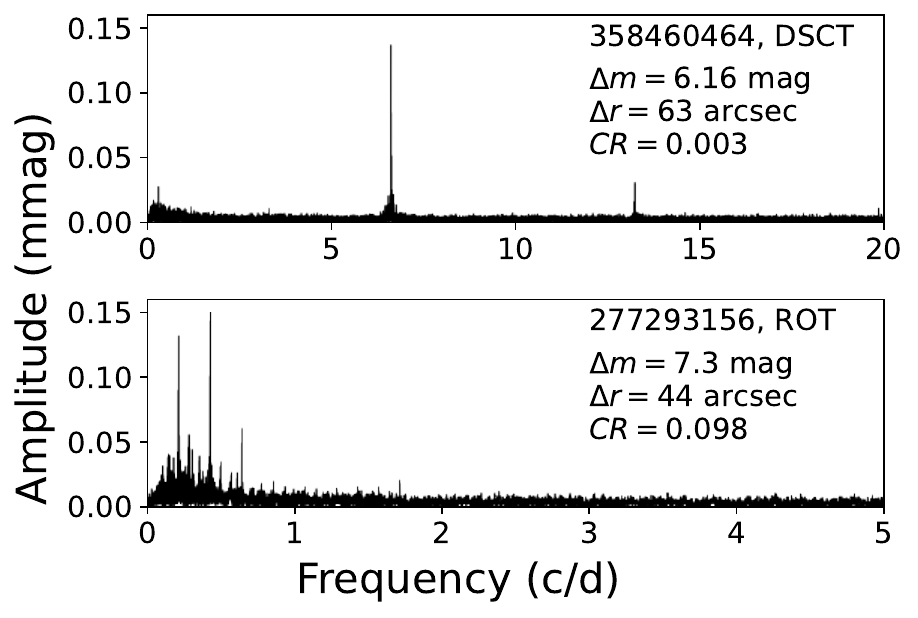}\\
\includegraphics[width=0.24\textwidth]{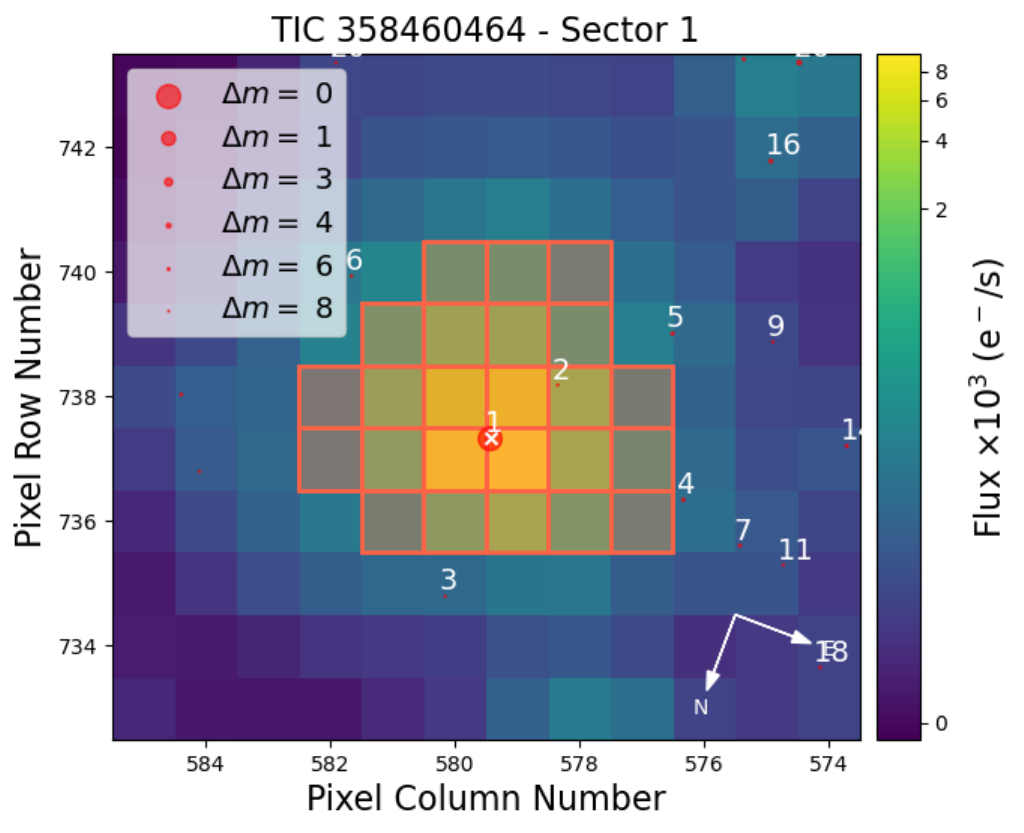}\includegraphics[width=0.24\textwidth]{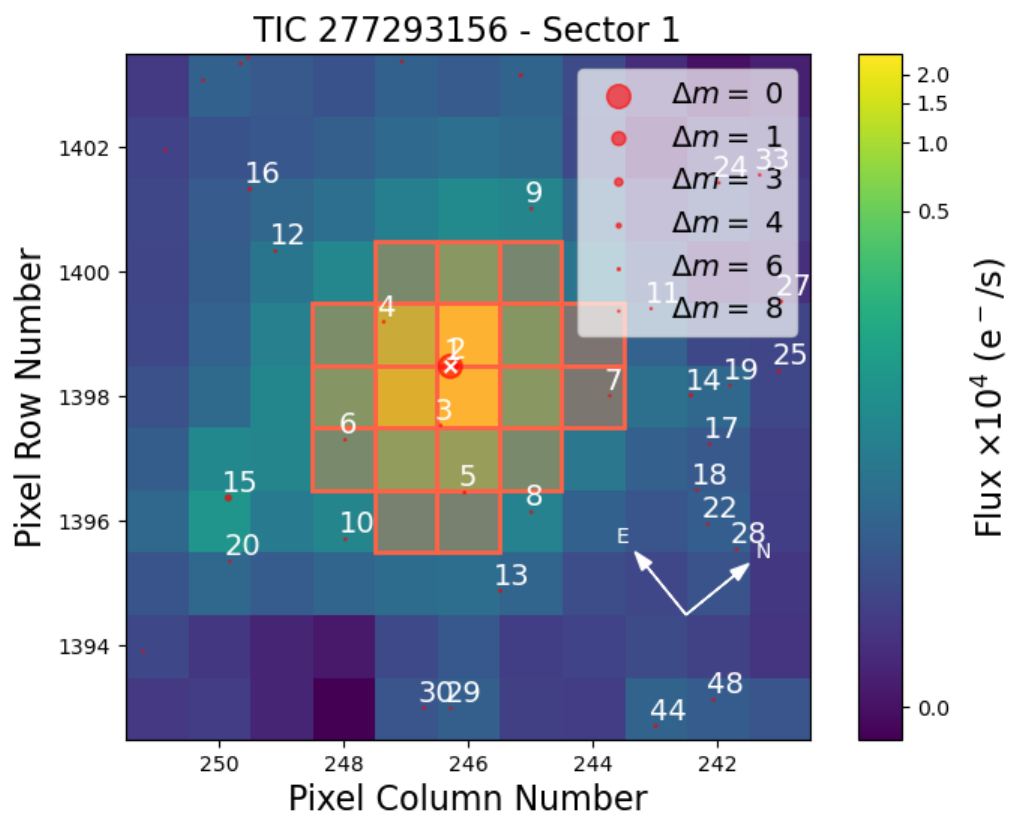}
\caption{Two identified variable stars that are actually not variable and their light is contaminated by nearby stars. The numbers show the TIC identification, variability type, magnitude difference ($\Delta m$) between the contaminant and target star, distance of the target from the contaminant ($\Delta r$) and the contamination ratio ({\it CR}). In the charts generated with \textsc{Tpfplotter} \citep{Aller2020} in the bottom part of the figure, the contaminating stars are numbers 4 and 6 in the left and right-hand panels, respectively.
}
\label{Fig:Contaminants}
\end{figure}

It is important to note that there may be nearby large-amplitude variables, such as eclipsing binaries or Cepheids, located just a few pixels away from the target star. Even when the brightness difference is quite large (more than 5\,mag), these variables can contaminate the light of the target and cause false identification of variability. Two examples of this are TIC\,358460464 and TIC\,277293156, as shown in Fig.~\ref{Fig:Contaminants}. The contaminants are approximately 3 and 2 pixels away from TIC\,358460464 and TIC\,277293156, respectively. Additionally, the contaminants are 6.2 and 7.3\,mag fainter than their respective target stars.

The contaminant of TIC\,358460464 is an EW ASAS-SN\,J075559.67-644238.3 which is a 15.2-mag star. Only the second and third harmonics of the orbital frequency are apparent in the frequency spectrum of TIC\,358460464, thus, we classified it as DSCT (top panel of Fig.~\ref{Fig:Contaminants}). On the other hand, three harmonics of the pulsation frequency of ASAS-SN\,J053737.66-702553.4 (DCEP star, 15.62\,mag) are apparent in the frequency spectrum of TIC\,277293156 (bottom panel of Fig.~\ref{Fig:Contaminants}), which makes it to belong to the ROT category. 

In both cases, the classification is wrong, although the contamination ratio is as low as 0.34\,\% in TIC\,358460464 (in TIC\,277293156 it is 10\,\%). Such wrongly identified stars add noise to the statistics when investigating common characteristics of the stars in particular variability classes. After this finding, we cross-matched our sample with {\it Gaia} DR3 variable star catalogue \citep{Rimoldini2023} and ASAS-SN variable star catalogue \citep{Jayasinghe2018} within a 100 arcseconds (5 px) radius and compared the frequencies (and their harmonics) from the catalogues with frequencies identified by us. This helped us to identify 12 contaminants (Table~\ref{Tab:Contaminants}), which are typically eclipsing binaries and pulsating stars. We have therefore classified our contaminated stars mostly as eclipsing binaries, DSCT and VAR. Such stars are marked with an additional ‘-C’ in the column type in Table~\ref{Tab:Main}. 

\setlength{\tabcolsep}{4pt}
\begin{table*}
\tiny
\caption{Identification of the contaminating stars.}             
\label{Tab:Contaminants}      
\centering                          
\begin{tabular}{c c c c c c c c c c}        
\hline\hline                 
TIC	&	Type	&	G (mag)	&	Type-Cont	&	ID ASAS-SN	&	ID {\it Gaia} DR3	&	$\Delta r$ (arcsec)	&	$\Delta m$ (mag)	&	$f_{\rm TESS}$ (c/d)	&	$f_{\rm Literature}$ (c/d)	\\ \hline
30469454	&	VAR	&	9.7	&	EW	&	J045847.67-661118.7	&		&	85.0	&	4.1	&	0.7569	&	1.5138	\\
141484981	&	DSCT|EW	&	9.3	&	EW	&	J054857.15-733629.1	&	4650219988274730000	&	62.4	&	5.0	&	2.9631	&	5.9262	\\
177022232	&	DSCT|EW	&	8.8	&	EW	&	J064932.91-674043.5	&	5281603511044130000	&	64.7	&	7.0	&	3.6288	&	7.2574	\\
260127863	&	DSCT+LF	&	9.9	&	EW	&	J060839.75-582809.0	&	5494885264769510000	&	75.8	&	4.5	&	3.5486	&	7.0970	\\
260160453	&	VAR	&	7.9	&	ECL	&		&	5482506511890750000	&	67.7	&	5.4	&	1.6843	&	3.3685	\\
308538268	&	VAR	&	9.0	&	EW	&	J080338.45-603047.4	&	5290787834752610000	&	77.4	&	4.7	&	2.2401	&	4.4810	\\
340312485	&	VAR	&	8.9	&	EW	&	J073934.70-595741.0	&		&	50.2	&	5.8	&	4.4852	&	8.9682	\\
349647697	&	GDOR	&	8.8	&	RRC	&	J072849.42-643817.6	&		&	78.5	&	7.5	&	3.6138	&	1.8951	\\
358460464	&	DSCT+LF	&	9.0	&	EW	&	J075559.67-644238.3	&	5275843994260360000	&	62.8	&	6.2	&	3.3086	&	6.6174	\\
358465355	&	EB	&	9.6	&	EB	&	J075605.94-601606.3	&	5290937849368120000	&	48.9	&	4.5	&	1.7528	&	3.5050	\\
372911993	&	DSCT+GDOR	&	9.0	&	ECL	&		&	5289793459627240000	&	51.3	&	5.6	&	2.8177	&	5.6355	\\
410452496	&	VAR	&	9.6	&	EW	&	J075944.49-612807.4	&	5289891938931840000	&	85.9	&	4.5	&	2.3923	&	4.7846	\\
\hline \hline                                   
\end{tabular}
\tablefoot{The column ‘Type-Cont’ gives the type of the contaminant from literature. If both ASAS-SN and {\it Gaia} DR3 designations are given, the contaminant was known to both ASAS-SN and {\it Gaia} DR3 variable stars catalogues \citep{Jayasinghe2018,Rimoldini2023}.}
\end{table*}

This exercise underscores the risks of misidentifying and misclassifying variable stars, which is far more common than we previously realized. In the {\it Gaia} DR3 variable star catalogue, only 20\,\% of stars in the vicinity of our target stars have had their frequencies determined. Identifying the true variable star is further complicated by the fact that there may be multiple stars within 100 arcseconds of the target (we only considered the brightest one). Additionally, bright stars can contaminate the light from even larger distances than 5 pixels. As a result, we can expect that the number of contaminants is much larger than the 12 cases we list in Table~\ref{Tab:Contaminants}.

Finally, there is high probability that many of the sample stars are bound in physical binary or multiple systems. In such cases, the observed variability is a combination of variability of system components. This can also lead to miss-classification of the variability type.  

\section{Results and discussion}\label{Sect:Results}

We have found that 50.9\,\% of the 1171 stars in our sample (excluding 33 duplicates) show clear variability. This is in agreement with the percentage of variable stars in the northern TESS CVZ (51\,\%, S22). We were able to determine the type of variability in 67.2\,\% of the stars, which is a better performance than in S22, where only 60\,\% of stars were unambiguously classified. Remaining 372 variable stars were unclassified\footnote{Considering ROTM|ELL stars as categorized stars.}. The higher number of data points, longer time span, and more sectors observed may have contributed to our better performance in classifying the stars.

\begin{figure}
\centering
\includegraphics[width=0.48\textwidth]{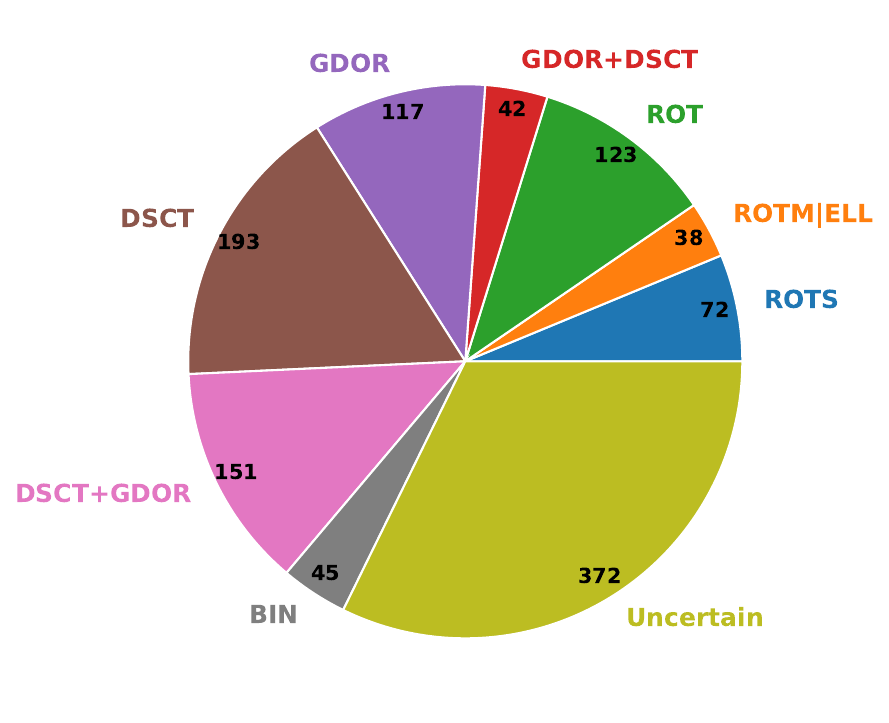}
\caption{Types of stars identified in our sample.
}
\label{Fig:Types_Stats}
\end{figure}

The numbers of the classified variable stars are shown in Fig.~\ref{Fig:Types_Stats}. We have identified a total of 503 pulsating stars, which includes DSCT, GDOR and their hybrids plus one RRAB star (43\,\% of the variable stars). This count is about 22\,\% less than that of the northern TESS CVZ (S22). We can only speculate about the reasons for the discrepancy.

\begin{figure}
\centering
\includegraphics[width=0.48\textwidth]{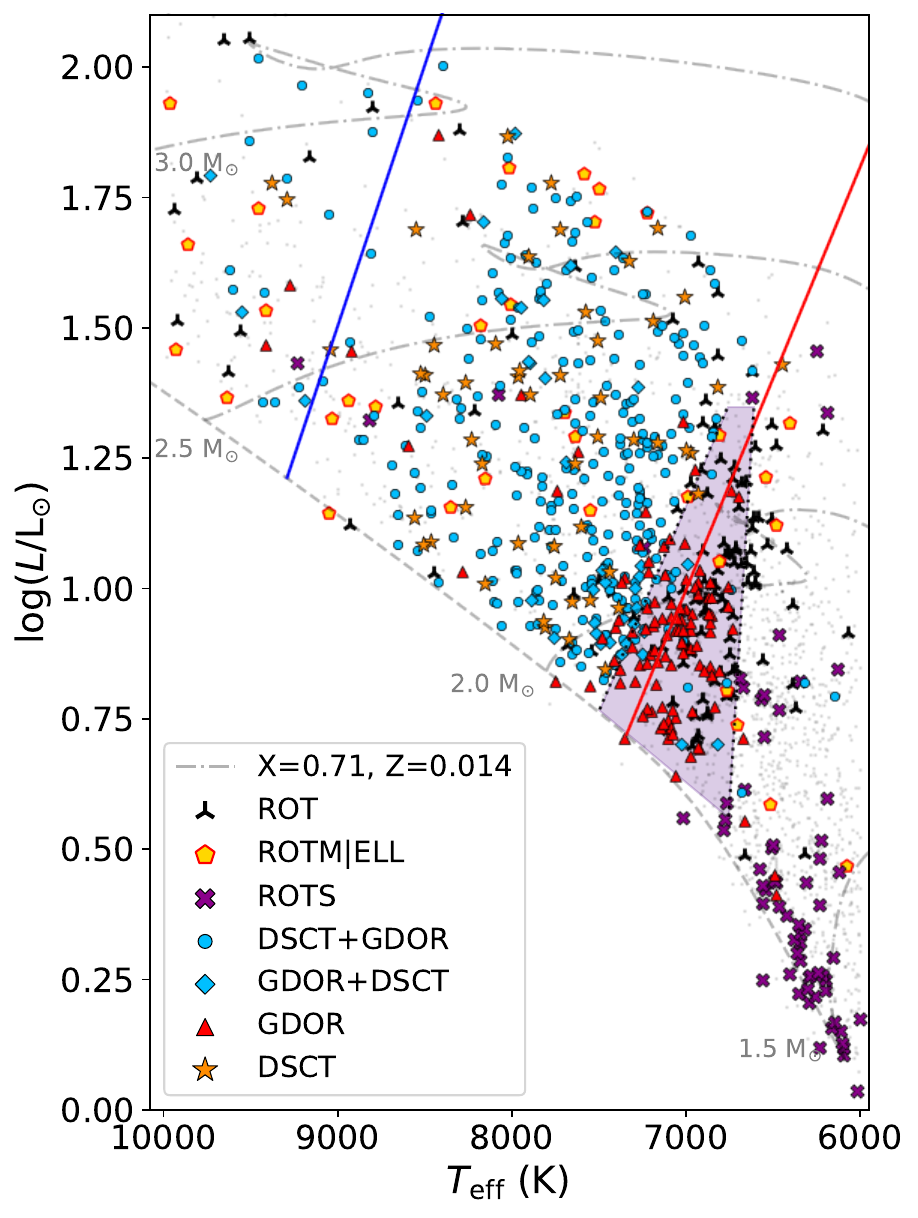}
\caption{Hertzsprung-Russell diagram showing our categorized stars. The dashed line shows the zero-age main sequence, while the evolutionary tracks are shown with the dash-dotted lines. The empirical boundaries of the DSCT instability strip are shown with blue and red continuous lines. The ZAMS, evolutionary tracks and instability strip boundaries are taken from \citet{Murphy2019}. The GDOR instability region \citep{Dupret2005} is shown with the shaded area limited by the dotted lines. All the stars in our sample including non-variable stars are shown with the grey dots.
}
\label{Fig:HRD}
\end{figure}

The diagram in Fig.~\ref{Fig:HRD} shows that DSCT and hybrid stars are evenly distributed throughout the empirical instability strip from \citet{Murphy2019}. Pure GDOR stars, on the other hand, are clustered within the theoretical GDOR instability strip from \citet{Dupret2005}, although they can also be found across the entire temperature range. At the hotter end of the scale, GDOR and DSCT stars are likely to mix with SPB and $\beta$\,Cep stars.

As expected, the ROTS stars are found at the lowest temperatures between 6000-7000\,K, while ROTM|ELL stars are spread out across the entire temperature range. This indicates that the population of ROTM|ELL is a mix of ELL and ROTM stars. The ROT stars tend to prefer temperatures around the cooler GDOR edge but can also be found spread out up to 10000\,K, similar to the distribution found in S22.

We cross-matched our sample with other catalogues within the distance of 21" (1 px) from our targets to compare the classifications. We used \textsc{Topcat} \citep{Taylor2005} and CDS X-Match service \citep{Boch2012,Pineau2020} to cross-match the catalogues. 
We can use two parameters describing the agreement of the two catalogues. The first parameter is the Classification Agreement Percentage (CAP), which gives the percentage of stars with the same classification in both catalogues. The second parameter is the Detection Agreement Percentage (DAP), which gives the fraction of stars identified and classified as VAR and other variability types in our sample while being assigned with the same variability type or as VAR in the other catalogue. 

For instance, if we classify a star as DSCT and the catalogue identifies it as VAR (and vice versa), it will count towards DAP. However, if we classify a star as DSCT and the catalogue identifies it as EW, it will not count. Similarly, if the catalogue gives a variability type, and we do not detect any variability, it will not count. DAP is useful in reflecting the uncertainty in the classification, as it gives a chance that the classification is correct in at least one of the compared catalogues.

There are 97 sample stars present in the VSX catalogue \citep{Watson2006}. This catalogue is a compilation of variable stars identified in various sky surveys and projects but also by individual observers (for example, NSVS \citep{Wozniak2004}, ZTF \citep{Chen2020}, CzeV catalogue \citep{Skarka2017}). However, only 85 of these 97 stars have been categorized. In 15 of the VSX-declared variables, we did not notice any variation. Out of the remaining 82 stars that have been classified in the VSX catalogue, we agree with the classification of only 26 stars (32\,\%). The CAP for our classification and VSX is, therefore, 26 out of 97 (27\,\%). 

\begin{figure}
\centering
\includegraphics[width=0.48\textwidth]{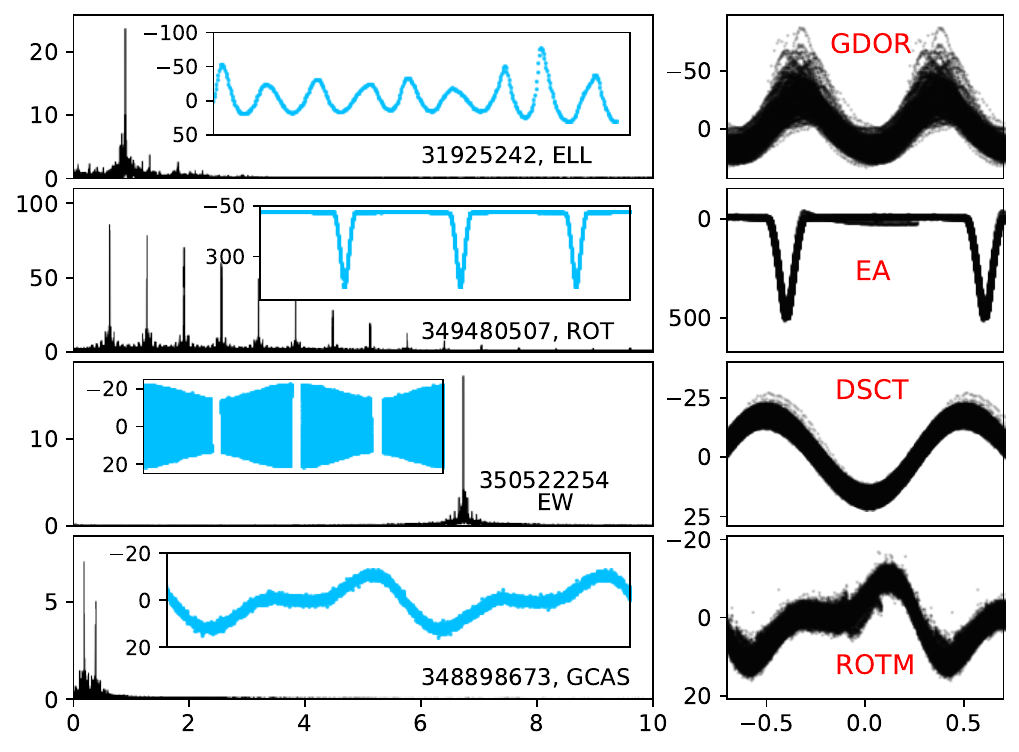}
\caption{Stars with wrong classification in the catalogues. The left panels show the frequency spectra with the part of the time series (the catalogue classification is given together with the TIC number), while the right-hand panels show the data phase-folded with the dominant frequency together with the correct variability type.
}
\label{Fig:Wrong}
\end{figure}

We show examples of where our classification differs from that of the VSX catalogue in Fig.~\ref{Fig:Wrong}. Specifically, we have classified TIC\,31925242 (CR Hyi) as a GDOR variable, whereas it is classified as ELL in the VSX catalogue (see the top panel of Fig.~\ref{Fig:Wrong}). Similarly, TIC 350522254 (XY Pic) is classified as DSCT by us\footnote{The variability type DSCT of XY Pic was already discussed by \citet{Dal2005}.}, while it is classified as EW in the VSX catalogue. The data of TIC\,350522254 show a beating caused by the presence of two close frequencies at about 6.7\,c/d (Fig.~\ref{Fig:Wrong}). For the VSX catalogue, the DAP 77\,\%.

Out of the 84 stars in our sample, 80 are classified as VAR in the ASAS-SN catalogue \citep{Jayasinghe2018,Jayasinghe2019}, while only four are categorized. Among the four categorized stars, one star is classified as GCAS (our ROTM|ELL, TIC\,348898673, see Fig.~\ref{Fig:Wrong}), and another star is listed as ROT (our EA, TIC\,349480507). However, we did not detect any variation in the remaining two stars that should be classified as GCAS according to \citet{Jayasinghe2018,Jayasinghe2019}. Furthermore, we did not detect variation in 9 of the declared variable stars. It is worth noting that there is no agreement between our classification and the ASAS-SN catalogue (CAP$=0$\,\%). The DAP for our classification and the ASAS-SN catalogue is 85\,\%, but this is only because most of the stars are labelled as VAR in the ASAS-SN catalogue.

We crossmatched our sample of stars also with the {\it Gaia} DR3 variables \citep{Rimoldini2023} within a 21" radius, and found 219 entries for our stars\footnote{By omitting three stars that are apparent blends with other stars in our sample.}. The CAP parameter is 48\,\% while DAP is 57\,\%. The high value of CAP is because most stars in our sample have a broad variability type classification in \citet{Rimoldini2023}. For example, {\it Gaia} DR3 classification ACV|CP|MCP|ROAM|ROAP|SXARI complies with our ROT, ROTS, ROTM|ELL stars, SOLAR LIKE complies with our ROTS, ROT and ROTM|ELL classes, DSCT|GDOR|SXPHE complies with our GDOR, DSCT and their hybrid classes.
We found no variability in 35 stars declared as variables in \citet{Rimoldini2023}.

In their study, \citet{Fetherolf2023} created a comprehensive list of variable stars from the TESS prime mission by searching for periodicity and variability using the Lomb-Scargle periodogram \citep{Lomb1976,Scargle1982}. Since we used the same data and analysis method, we expected to obtain similar results. Our sample overlapped with  1141 stars from \citet{Fetherolf2023} catalogue, including 16 duplicates and 9 false positives (stars that were blends with faint large-amplitude variables). Unfortunately, \citet{Fetherolf2023} did not provide a classification of the variable stars, so we could only calculate the DAP, which was only 63\,\%. Most of the stars that we disagreed on did not exhibit apparent variability and displayed low-frequency peaks similar to those shown in panel (c) of Fig.~\ref{Fig:VAR}. The frequencies of variable stars found by \citet{Fetherolf2023} perfectly matched with frequencies found by us (including our VAR stars). However, in stars that we marked as stable, the frequencies identified by \citep{Fetherolf2023} differed from the frequencies identified by us. This is an additional argument against assigning these stars as variable.

Fourty-four stars from our sample are listed in the binary-star catalogue by \citet{Prsa2022}. However, we classified three of them not as eclipsing binaries but as ROTM|ELL  (TIC 382512330, 306578324, and 141127435) and TIC\,142148228 and 150187916 as VAR stars. The CAP and DAP with \citet{Prsa2022} are both 89\,\%. This result is no surprise since variations of eclipsing binaries are characteristic and usually have large amplitudes.

Regarding rotational variable stars, 29 stars from our sample appear in \citet{Sikora2019}. The variations are connected with rotation or orbital motion only in 16 of them. The rotational nature of the 13 rest stars is ambiguous. The CAP and DAP are both 55\,\%. 

Twelve of our sample stars are listed in \citet{Antoci2019} which deals with DSCT stars. Nine of them were identified as DSCT (or hybrids) also by us, and one of the stars was found as constant in agreement with \citet{Antoci2019}. In TIC\,260416268, which was identified as rot/binary Am star by \citet{Antoci2019}, we detected no variability. The same applies to the roAp candidate TIC\,238185398. The CAP and DAP are both 83\,\%.

The discrepancies between our results and results from literature arise from the superior characteristics of the TESS data compared to other catalogues meaning better precision, cadence and often also the time span. The agreement in classification with previous studies and catalogues is summarized in Table~\ref{Tab:Agreement}. 

\begin{table}
\caption{Agreement (1st column) in per cent of our classification with the large catalogues and dedicated studies.}             
\label{Tab:Agreement}      
\centering                          
\begin{tabular}{c c c}        
\hline\hline                 
CAP/DAP (\%) & Type & Catalogue \\    
\hline                        
27/77 & General & VSX \citep{Watson2006}\\      
0/85 & General & ASAS-SN \citep{Jayasinghe2018} \\ 
48/57 & General &{\it Gaia} DR3 \citep{Rimoldini2023} \\ 
-/63 & General & TESS PM \citep{Fetherolf2023}\\
89/89 & Binaries & \citep{Prsa2022} \\
55/55 & ROT & \citep{Sikora2019a} \\
83/83 & DSCT & \citep{Antoci2019}\\
\hline
\end{tabular}
\end{table}

\section{Conclusions}\label{Sect:Conclusions}

We carefully classified variable stars using data from TESS from Cycles 1 and 3. This data is useful for detecting fast variations, such as DSCT pulsations, as well as variations that occur over several days. We discussed the limitations of our classifications, including the general shortcomings in variable stars classification. Furthermore, we compared our results with existing databases and variable star catalogues. 

We confirm the previous issues with the classification of rotationally variable stars and hours- to days-long pulsations of GDOR type. It is also often difficult to distinguish between ellipsoidal and/or contact binary variables and spotted stars without additional information. Additionally, the classification can be influenced by nearby stars, and the level of contamination depends on the distance of the contaminating star, the amplitude of variations, and the difference in brightness. This mainly affects the classification of rotationally variable stars and stars that show a combination of different types of variation.

We found intriguing discrepancies in the classification in variable stars catalogues that can be up to 100\,\%. To address this issue, we have introduced two parameters - CAP and DAP. CAP reflects the classification, while DAP concerns the detection of variability without a specific variability type. Our results show that we have a relatively good agreement with the {\it Gaia} catalogue \citep[CAP$=48$\,\%][]{Rimoldini2023}. The best agreement was with the studies dedicated to particular variability types. For instance, the agreement of our classification was almost 90\,\% with the database of eclipsing binaries \citep{Prsa2022} and 83\,\% in case of DSCT variables \citep{Antoci2019}. Surprisingly, the agreement of our results with a study by \citet{Fetherolf2023}, that was also based on the TESS data, was only 63\,\%. We have also demonstrated that large-amplitude variables can contaminate the light of the target stars up to 100 arcseconds from the star, which can lead to incorrect identification and classification.

It has been shown that classifying variable stars is a complex task and may produce ambiguous or incorrect results. It is not advisable to blindly use samples of variable stars without individually checking each star. This is especially important in usage of samples based on automatic classifiers. To create reliable catalogues of variable stars, it is necessary to analyze not only single-channel photometric observations but also multi-color and spectroscopic observations, and to estimate reddening and distance. Additionally, one must be very careful in handling contamination. Only then can we be confident that the variations of specific objects are caused by the proposed effect(s). 

The first {\it PLATO} mission field \citep[LOPS2,][]{Nascimbeni2022} will include the southern TESS CVZ. Our classification represents a good starting point for the studies of stellar variability in the {\it PLATO} field. 

\begin{acknowledgements}
      MS acknowledges the support by Inter-transfer grant no LTT-20015. 
      This paper includes data collected with the TESS mission. Funding for the TESS mission is provided by the NASA Explorer Program. Funding for the TESS Asteroseismic Science Operations Centre is provided by the Danish National Research Foundation (Grant agreement no.: DNRF106), ESA PRODEX (PEA 4000119301) and Stellar Astrophysics Centre (SAC) at Aarhus University. We thank the TESS team and staff and TASC/TASOC for their support of the present work. We also thank the TASC WG4 team for their contribution to the selection of targets for 2-min observations. The TESS data were obtained from the MAST data archive at the Space Telescope Science Institute (STScI). 
\end{acknowledgements}

\bibliographystyle{aa}
\bibliography{references}

\end{document}